\documentclass[11pt,a4paper]{article}

\usepackage[utf8]{inputenc}
\usepackage[T1]{fontenc}
\usepackage[english]{babel}
\usepackage{lmodern}
\usepackage{cmap}
\InputIfFileExists{glyphtounicode.tex}{}{}
\usepackage{microtype}
\usepackage{setspace}
\usepackage[margin=2.5cm]{geometry}
\usepackage{graphicx}
\usepackage{booktabs}
\usepackage{array}
\usepackage{amsmath,amssymb}
\usepackage{xcolor}
\usepackage{caption}
\usepackage[square,numbers,sort&compress]{natbib}
\usepackage{hyperref}
\hypersetup{
  unicode=true,
  colorlinks=true,
  linkcolor=blue,
  citecolor=blue,
  urlcolor=blue,
  pdftitle={endoExplain: A reproducible protocol for auditing score-localisation discordance in colonoscopy image classifiers},
  pdfauthor={Roberto Alcaraz Machado and Ana Lucía Rodríguez Blanco},
  pdfsubject={Reproducible score-localisation auditing in colonoscopy image classifiers},
  pdfkeywords={colonoscopy; explainable AI; class activation mapping; calibration; localisation audit; reproducibility}
}
\newcommand{\ManifestDigestShort}{6b12b62f7c72}

\newcommand{\PrimaryValAUPRC}{0.983}
\newcommand{\PrimaryTestAUPRC}{0.968}
\newcommand{\PrimaryTestAccuracy}{0.973}
\newcommand{\CalibrationTemperature}{0.829}
\newcommand{\RawTestECE}{0.0167}
\newcommand{\ScaledTestECE}{0.0115}
\newcommand{\RawTestNLL}{0.0838}
\newcommand{\ScaledTestNLL}{0.0829}
\newcommand{\RawTestBrier}{0.0205}
\newcommand{\ScaledTestBrier}{0.0206}
\newcommand{\PrimaryHighN}{172}
\newcommand{\GradCamDiscordance}{50.6\%}
\newcommand{\GradCamPpDiscordance}{62.2\%}
\newcommand{\XGradCamDiscordance}{54.7\%}
\newcommand{\HiResCamDiscordance}{5.8\%}
\newcommand{\EigenCamDiscordance}{4.1\%}

\title{endoExplain: A reproducible protocol for auditing
score--localisation discordance in colonoscopy image classifiers}
\author{
Roberto Alcaraz Machado$^{1,*}$\\
ORCID: 0009-0009-1551-7794
\and
Ana Luc\'ia Rodr\'iguez Blanco$^{2}$\\
ORCID: 0009-0009-6449-1797
}
\date{}

\begin{document}
\maketitle
\noindent $^{1}$Biomedical Engineer, Independent Researcher, Alicante, Spain.\\
$^{2}$Department of General and Digestive Surgery, Hospital Universitario de San Juan, Carretera Nacional 332 Alicante--Valencia, 03550 Sant Joan d'Alacant, Alicante, Spain.\\
$^{*}$Corresponding author: \texttt{eroberam@gmail.com}
\begin{abstract}
\singlespacing
\textbf{Background and objective:} A high classifier score and a plausible
class-activation map (CAM) are often presented together, although neither
establishes that the other is reliable. We introduce endoExplain as a
reproducible protocol for auditing score--localisation discordance rather than
as a new detector or explanation algorithm.
\textbf{Methods:} Content hashing separated HyperKvasir development images
from 1,000 masked images before training. EfficientNet-B0, ResNet-34 and
ConvNeXt-Tiny were trained with three seeds each. Scores were temperature
scaled using validation data only. Grad-CAM, Grad-CAM++, XGrad-CAM, HiResCAM
and Eigen-CAM were evaluated on identical image--mask pairs, alongside random
and centre baselines. Outcomes combined peak localisation, overlap, a top-20\%
deletion response, score-threshold sensitivity and adjustment for lesion size
and centrality. The selected checkpoint was transferred without retraining to
three external mask cohorts.
\textbf{Results:} Temperature scaling reduced test expected calibration error
from \RawTestECE{} to \ScaledTestECE{}. Among \PrimaryHighN{} reserved images
with scaled score at least 0.90, peak-outside-lesion rates ranged from
\EigenCamDiscordance{} to \GradCamPpDiscordance{} across CAMs. Method
dependence remained evident across architectures and seeds, although method
rankings were not universal. Spatial alignment and deletion response were not
interchangeable. A random-deletion control also produced positive logit drops,
limiting specificity claims based on deletion alone. External positive-mask
results were dataset dependent. A
source-category audit also exposed that 149/155 test
positives were dyed-lifted polyps, materially bounding classifier claims.
\textbf{Conclusions:} endoExplain makes calibration, spatial agreement,
perturbation response and transfer separately inspectable. The results caution
against using a score or visually persuasive CAM as evidence of lesion
localisation or model reasoning.
\end{abstract}


\section{Introduction}
\label{sec:introduction}

Artificial intelligence (AI) systems for colonoscopy are increasingly
evaluated in clinically relevant settings, and computer-aided detection can
increase adenoma and polyp detection in controlled studies
\citep{wang2018realtime,zha2024umbrella,park2024multicenter}. Alongside a
prediction, many systems display a heatmap and a scalar score. The combination
is visually compelling: the number appears to quantify certainty and the map
appears to show where the evidence lies. That interpretation is stronger than
either output warrants.

A classifier score quantifies a model output for a specified target; it is not
automatically a calibrated probability. A CAM is a transformation of internal
activations or gradients; it is not automatically a lesion segmentation or a
causal account. Calibration, localisation and faithfulness therefore answer
different questions. A classifier can assign a high score while its CAM peaks
outside the annotated lesion. A map can overlap a lesion because both are
central or large, even if deleting the highlighted pixels does not reduce the
target logit. Conversely, a low-overlap map does not by itself invalidate the
class prediction. These distinctions matter whenever score and heatmap are
presented as mutually reinforcing evidence.

Grad-CAM and its extensions are widely used because they are inexpensive and
model compatible \citep{selvaraju2017gradcam,chattopadhay2018gradcampp,
fu2020xgradcam,draelos2021hirescam}. Yet saliency methods can respond to model
and label randomisation in unexpected ways, and visually plausible maps may
not be faithful \citep{adebayo2018sanity,samek2017evaluating,hooker2019roar,
rong2022road}. Medical-imaging evaluations likewise show that explanation
quality depends on the metric and task \citep{wickstrom2020xai,
arun2021assessing}. A robust audit must therefore control the image, class,
model checkpoint and map threshold; compare methods by image; include spatial
baselines; and distinguish lesion agreement from perturbation response.

Two additional sources of optimism are common. First, near-duplicate images
can cross development and explanation pools when public datasets are joined by
filename rather than content. Second, a high raw score is often described as
``confidence'' without calibration or subgroup inspection. Both problems can
produce a coherent-looking narrative even when the evaluated target differs
from the intended clinical construct. Public polyp masks make spatial auditing
possible \citep{borgli2020hyperkvasir,silva2014embedded,ali2023polypgen,
jha2024polypdb}, but do
not remove these design requirements.

We address this gap with endoExplain, a computational protocol and reference
implementation for case-level score--localisation audits. It does not propose
a new CAM, classifier or clinical device. Its methodological contribution is
the coordinated enforcement of: (i) content-defined study roles; (ii)
validation-only model selection and calibration analysis; (iii) paired
multi-method analysis; (iv) localisation and deletion outcomes kept separate;
(v) architecture, seed, threshold, lesion-size and centrality sensitivity; and
(vi) frozen external transfer with traceable artefacts.

The study asks four ordered questions. \textbf{RQ1:} Among images assigned a
high temperature-scaled target score, how often does each CAM peak outside the
lesion?
\textbf{RQ2:} Does that result persist across score thresholds, architectures,
training seeds, lesion size and lesion centrality? \textbf{RQ3:} Do lesion
alignment and deletion response support the same interpretation? \textbf{RQ4:}
How does a frozen audit change on external positive-mask datasets? The ordering
provides the manuscript's thread: model discrimination is characterised first,
then the evidence displayed for individual predictions is audited, then its
robustness and transfer are tested.

\section{Related work and methodological position}
\label{sec:related}

CAM methods differ in the signal used to weight convolutional features.
Grad-CAM pools class gradients; Grad-CAM++ changes the gradient weighting;
XGrad-CAM uses an axiomatic weighting; and HiResCAM retains the elementwise
product between activations and gradients before summing channels
\citep{selvaraju2017gradcam,chattopadhay2018gradcampp,fu2020xgradcam,
draelos2021hirescam}. HiResCAM provides a faithfulness result for specified CNN
architectures, but this architectural property does not make a CAM a lesion
mask. Eigen-CAM instead uses principal activation structure and is not
class-specific \citep{muhammad2020eigencam}; it is retained here as an
activation comparator and marked with an asterisk throughout.

Explanation evaluation frameworks organise localisation, faithfulness,
robustness and complexity metrics, illustrating that no single score captures
all desirable properties \citep{hedstrom2023quantus}. Perturbation methods
such as deletion, ROAR and ROAD ask whether model output changes when attributed
regions are removed or corrupted \citep{petsiuk2018rise,hooker2019roar,
rong2022road}. These tests are informative but themselves create altered,
possibly out-of-distribution images. We therefore call our deletion outcome a
``perturbation response'' and do not equate it with causal proof.

In medical and endoscopic imaging, quantitative work has compared saliency
maps with expert or lesion annotations and documented instability across
methods \citep{wickstrom2020xai,arun2021assessing}. endoExplain builds on that
literature by centring the sampling question: what does spatial performance
look like specifically among images above a predeclared threshold on a
validation-fitted, temperature-scaled target score? Table~\ref{tab:positioning}
distinguishes the primary scope of
representative prior work from the integrated protocol evaluated here. A dash
does not mean that a method cannot support a feature; it means the feature was
not the central contribution of the cited work.

Recent work sharpens the translational gap. A clinician-informed study of
concept-based explanations in gastrointestinal disease detection found that
intuitive displays alone did not establish clinical readiness
\citep{storas2025clinical}. A 2026 medical-imaging assessment framework
separates consistency, plausibility, fidelity and usefulness rather than
treating ``explainability'' as one outcome \citep{lago2026evaluating}.
Endoscopic-AI quality guidance likewise calls for system-level evaluation
beyond model accuracy \citep{mori2025quality}. These distinctions motivate our
separate reporting of spatial agreement, perturbation response and intended
use.

\begin{table}[!htbp]
\centering
\caption{Methodological position of endoExplain. Columns denote features
treated as central study components, not a ranking of the cited methods.}
\label{tab:positioning}
\footnotesize
\begin{tabular}{>{\raggedright\arraybackslash}p{0.25\linewidth}
                *{5}{>{\centering\arraybackslash}p{0.115\linewidth}}}
\toprule
Work / primary scope & Mask alignment & Faithfulness & Temperature-scaled score conditioning & Model repeats & Frozen external audit \\
\midrule
Grad-CAM family \citep{selvaraju2017gradcam,chattopadhay2018gradcampp,fu2020xgradcam} & partial & -- & -- & -- & -- \\
HiResCAM \citep{draelos2021hirescam} & partial & architectural & -- & -- & -- \\
Medical saliency assessment \citep{wickstrom2020xai,arun2021assessing} & yes & partial & -- & partial & partial \\
Quantus \citep{hedstrom2023quantus} & metric & metric & -- & configurable & configurable \\
endoExplain (this study) & paired & deletion response & yes & 3 $\times$ 3 & yes \\
\bottomrule
\end{tabular}
\end{table}

\section{Materials and methods}
\label{sec:methods}

\subsection{Study design and data roles}

\begin{figure}[!htbp]
    \centering
    \includegraphics[width=\linewidth]{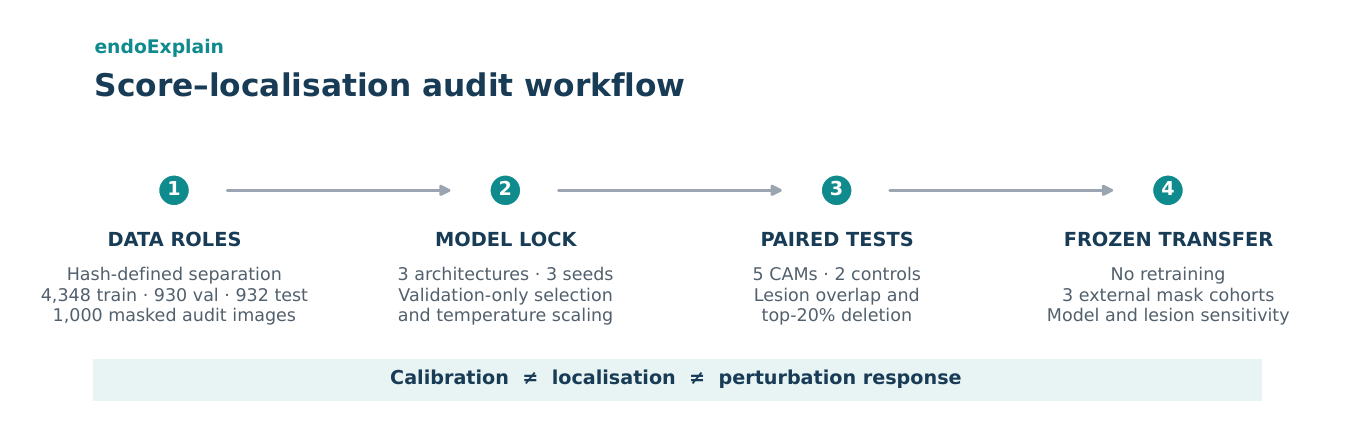}
    \caption{endoExplain workflow. Content-defined roles precede model
    development. Calibration uses validation data only. Localisation and
    perturbation outcomes are calculated on paired maps, followed by robustness
    analyses and transfer without retraining.}
    \label{fig:study_design}
\end{figure}

Figure~\ref{fig:study_design} summarises the protocol. All choices that could
use class labels---split construction, checkpoint selection and temperature
fitting---were completed before the spatial audit. Every CAM received the same
image, binary mask, target class, checkpoint and top-area rule. The unit of
analysis was the image.

HyperKvasir supplied labelled lower-gastrointestinal images and a segmented
polyp subset \citep{borgli2020hyperkvasir}. SHA-256 comparison established
that all 1,000 segmented images were byte-identical to files in the labelled
\texttt{polyps} class. They were assigned the immutable role
\texttt{spatial\_audit} before training and excluded from train, validation
and test. A $16\times16$ average hash added a conservative
appearance-collision rule based on exact 256-bit fingerprint equality (Hamming
distance 0), not approximate nearest-neighbour matching. No byte-identical or
identical-average-hash group crossed study roles in the final 7,210-row manifest (digest
\texttt{\ManifestDigestShort{}...}).

The binary target, \texttt{polyp\_family}, combined the \texttt{polyps} and
\texttt{dyed-lifted-polyps} source categories. Negatives were lower-GI anatomical,
bowel-preparation and inflammatory classes; upper-GI images were excluded as
clinically trivial negatives. Table~\ref{tab:composition} reports the source
composition rather than only the collapsed binary counts. Standard polyps were
rare in development: 19/721, 3/154 and 6/155 positives in train, validation and
test, respectively. Thus the classifier experiment mainly distinguishes
dyed-lifted polyps from heterogeneous lower-GI negatives. The 1,000 standard
polyp masks form a content-disjoint but visibly shifted spatial audit, not an
independent identically distributed classifier test.

\begin{table}[!htbp]
\centering
\caption{HyperKvasir roles and positive-source composition. The imbalance
between standard and dyed-lifted polyps bounds interpretation of classifier
performance.}
\label{tab:composition}
\small
\begin{tabular}{lrrrr}
\toprule
Role & Total & Negative & Dyed-lifted polyp & Standard polyp \\
\midrule
Training & 4,348 & 3,627 & 702 & 19 \\
Validation & 930 & 776 & 151 & 3 \\
Test & 932 & 777 & 149 & 6 \\
Spatial audit & 1,000 & 0 & 0 & 1,000 \\
\bottomrule
\end{tabular}
\end{table}

ETIS-LaribPolypDB supplied 196 rows \citep{silva2014embedded}; three
byte-identical frame groups were collapsed by mask union into 193 unique cases.
PolypGen supplied 1,537 single frames from six centres
\citep{ali2023polypgen}; two duplicates were collapsed and 125 empty-mask
frames were excluded from localisation, leaving 1,410 visible-polyp cases.
PolypDB supplied 3,934 unique still images from three centres and five
modalities \citep{jha2024polypdb}. Only its centre-wise tree was indexed
because the modality-wise tree reorganises the same collection.

Hash audits found no exact or average-hash overlap with the complete
HyperKvasir manifest or between external cohorts. The locally downloaded
LDPolypVideo subset contained 103 video-level-labelled AVI files but none of
the frame annotations described for the published benchmark
\citep{ma2021ldpolypvideo}; it was therefore reserved for a separate temporal
protocol. Because the analysed spatial subsets contain visible masked polyps
without compatible negative controls, they support spatial transfer only, not
external AUPRC, specificity or calibration.

\subsection{Classifiers, selection and calibration}

EfficientNet-B0, ResNet-34 and ConvNeXt-Tiny
\citep{tan2019efficientnet,he2016resnet,liu2022convnext} were trained with seeds
1129, 2089 and 3253 on the identical manifest. ImageNet-pretrained networks
received deterministic $256\times256$ validation/test images and training
images augmented by flips and mild brightness/contrast changes. Optimisation
used cross-entropy with label smoothing 0.03, AdamW, cosine learning-rate
decay, mixed precision and early stopping. The implementation used PyTorch,
Torchvision and Albumentations \citep{paszke2019pytorch,marcel2010torchvision,
buslaev2020albumentations,micikevicius2018mixed}. The three EfficientNet-B0 runs were
eligible for the prespecified primary checkpoint because that compact family
was chosen for the operational audit pipeline; validation AUPRC was the sole
selector. ResNet-34 and ConvNeXt-Tiny provided architecture sensitivity and
did not compete for primary selection. A shuffled-training-label run checked
the learning path. Test metrics were exported only after selection.

Raw softmax output was not treated as a calibrated probability. For each of the nine
checkpoints, scalar temperature $T>0$ was fitted by minimising binary negative
log-likelihood (NLL) on the validation split only \citep{guo2017calibration}.
If $z_i$ denotes the binary logit difference, the scaled score was
\begin{equation}
    \hat p_i = \sigma(z_i/T).
\end{equation}
The same fixed $T$ was applied to test and all spatial datasets. We report NLL,
Brier score and expected calibration error (ECE; ten equal-width bins), with a
reliability diagram. Scaling is monotonic and therefore does not change AUPRC
or the 0.50 decision boundary, but it changes which cases meet score thresholds
above 0.50. Because validation contained only three standard polyps, the scaled
score remains a property of the mixed development target and is not claimed as
a clinically calibrated standard-polyp probability.

\subsection{CAM generation and spatial outcomes}

The five evaluated maps were Grad-CAM, Grad-CAM++, XGrad-CAM, HiResCAM and
Eigen-CAM. Gradient methods targeted the positive-class logit. HiResCAM was
implemented through the independently maintained \texttt{pytorch-grad-cam}
library and checked against the authors' elementwise gradient--activation
definition; no code was copied from the local reference repository. Eigen-CAM
is class-agnostic and is marked with an asterisk. Seeded random heatmaps and a
centred Gaussian were spatial baselines. CAM deletion response was complemented
by an independent exact-area random-deletion control; no deletion outcome was
assigned to the spatial-baseline maps themselves.

Maps from the final convolutional block were resized to the image, normalised
per image to $[0,1]$ and compared with the binary lesion mask. Let $s_i$ be the
temperature-scaled positive score, $H_{im}$ the map for image $i$ and method
$m$, and $M_i$ its mask. The primary conditional failure indicator was
\begin{equation}
 F_{im}(c)=\mathbf{1}(s_i\geq c)\,
 \mathbf{1}\!\left[\arg\max H_{im}\notin M_i\right],
 \qquad c=0.90,
\label{eq:failure}
\end{equation}
and the reported rate conditions on $s_i\geq c$. This pointing-game endpoint
does not depend on map binarisation. Secondary outcomes were mean IoU after
retaining the top 20\% of map pixels, IoU below 0.10, attribution energy inside
the lesion, centre-of-mass distance, lesion area fraction and normalised
distance from lesion centroid to frame centre. For complete-population
summaries, unconditional pointing failure was additionally calculated across
all 1,000 spatial-audit images, irrespective of target score. Rank thresholding
retained exactly the requested number of pixels. When a top-area boundary was tied, the
remaining quota was selected by a fixed multiplicative permutation of flattened
pixel indices. Pointing game used the first row-major maximum. Constant maps,
maximum multiplicity and boundary-tie counts were exported: among 6,000
non-centre maps, one HiResCAM map was constant, and partial boundary ties
occurred in 1,233 maps. These deterministic conventions can affect tied pixels and are therefore
reported rather than assumed to be bias free.

\subsection{Perturbation response and robustness analyses}

For an empirical faithfulness stress test, the top 20\% of attributed pixels
were set to zero in normalised tensor space, corresponding to the ImageNet mean
RGB value before normalisation. We recorded the
change in positive logit and score between the original and deleted image. A
positive logit drop indicates that the highlighted region supported the target
under this intervention; zero or negative change does not. Since mean-value
deletion can create out-of-distribution boundaries, this outcome is analysed
alongside, not above, lesion alignment. As a perturbation control, exact-area
random deletion was repeated 20 times per image with deterministic
image--repetition seeds; confidence intervals resampled the 1,000 image-level
means. The centred Gaussian was retained only as a spatial control.

All seven maps were recomputed for every architecture--seed checkpoint on all
1,000 reserved images. Score-threshold sensitivity used $c=0.80$, 0.85, 0.90
and 0.95 and reported the number of eligible images at each point. Checkpoint
sensitivity was reported both on each checkpoint's own score-defined subset
and on the fixed 172 images selected by the primary checkpoint. Lesion area
was divided into dataset-specific tertiles. For the primary checkpoint, a
logistic model of pointing failure among cases with $s_i\geq0.90$ included CAM
method indicators (Grad-CAM reference), lesion area per 0.10 frame fraction,
lesion-centre distance per 0.10 and score per 0.10. A near-unpenalised ridge
term stabilised separation; image-cluster bootstrap intervals used 500
replicates. This model is an adjustment analysis, not a causal effect estimate.

\subsection{Statistics, qualitative cases and reproducibility}

Methods were paired by image. Primary-checkpoint CAM comparisons used exact
paired binomial tests for pointing failure and paired Wilcoxon tests for IoU,
with Holm correction within outcome. Percentile intervals resampled images
with seed 20260721; 2,000 replicates were used for the confirmatory spatial
summaries and 5,000 for classifier metrics and figure-specific mean IoU.
Patient, sequence and acquisition-centre identifiers were unavailable, so the
image bootstrap may understate uncertainty when frames are correlated.

Qualitative panels were selected only after the complete paired table was
created. The discordant example maximised method balance, then the across-CAM
IoU range and score, among high-score cases with at least one peak inside and
one outside the mask. The consensus example maximised mean IoU among
high-score cases whose five CAM peaks were all inside. Sample identifier broke
ties. This rule prevents manual selection by visual appeal.

All inputs and outputs carry dataset, sample, model, seed, target, map method,
threshold and calibration metadata. Scripts reject duplicate method--sample
rows, unpaired image sets and mixed map thresholds. The repository includes
the manifest builder, registered training matrix, calibration and robustness
runners, statistical audit, tests and release builder.

\section{Results}
\label{sec:results}

\subsection{Classifier selection, calibration and source-category audit}

The content audit reserved all 1,000 masked images and found zero
cross-role byte-identical or identical-average-hash groups. EfficientNet-B0 seed 2089 was selected
by validation AUPRC \PrimaryValAUPRC{}. Its test AUPRC was
\PrimaryTestAUPRC{} and accuracy at 0.50 was \PrimaryTestAccuracy{}. The
shuffled-label control remained near the positive-prevalence AUPRC baseline,
arguing against a broken export path.

\begin{figure}[!htbp]
    \centering
    \includegraphics[width=\linewidth]{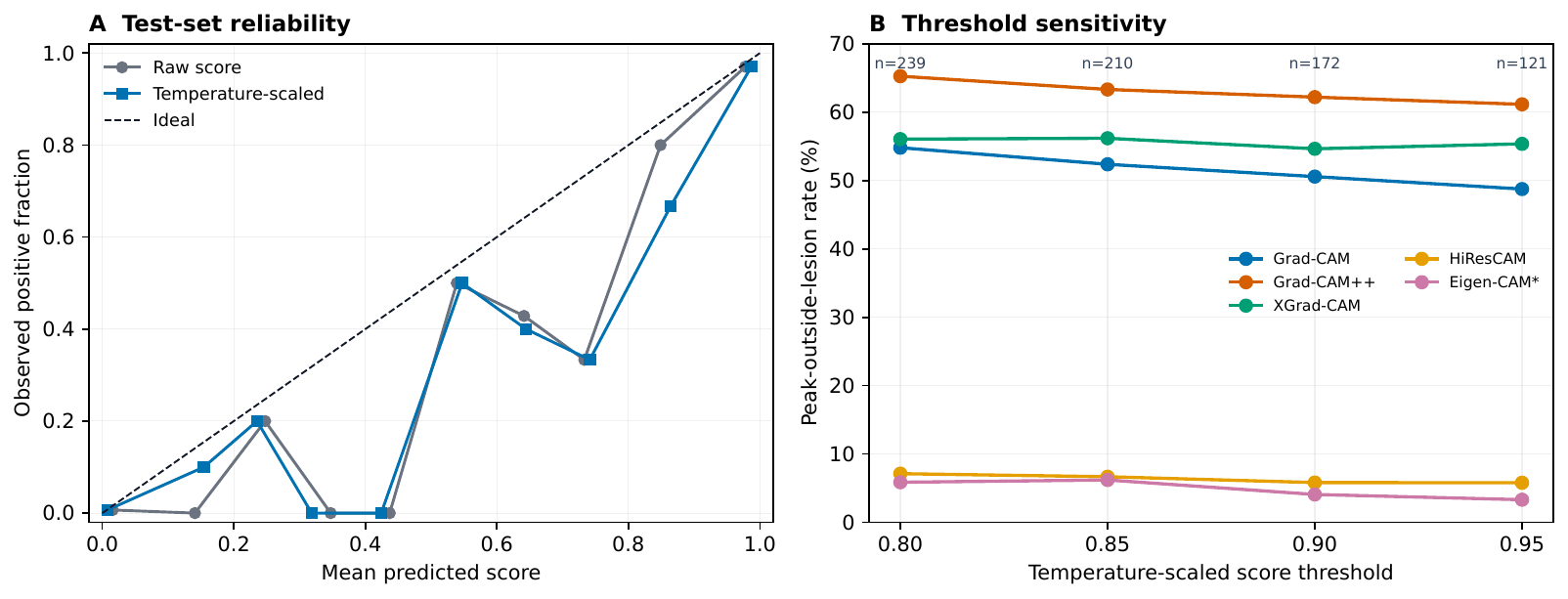}
    \caption{Calibration and score-threshold sensitivity. (A) Reliability of
    raw and validation-scaled scores on the held-out classifier test split.
    (B) Peak-outside-lesion rate on the reserved mask pool as the scaled-score
    threshold changes; labels give the number of eligible images. Eigen-CAM is
    class-agnostic and marked with an asterisk.}
    \label{fig:calibration}
\end{figure}

Validation-only fitting produced $T=\CalibrationTemperature{}$. On the test
split, NLL changed from \RawTestNLL{} to \ScaledTestNLL{} and ECE from
\RawTestECE{} to \ScaledTestECE{}, whereas Brier score changed from
\RawTestBrier{} to \ScaledTestBrier{} (Table~\ref{tab:calibration}). Thus
scaling improved ECE and slightly improved NLL but did not improve every proper
score. Figure~\ref{fig:calibration} shows that reliability estimates are also
sparse in intermediate bins.

\begin{table}[!htbp]
\centering
\caption{Calibration of the validation-selected EfficientNet-B0. Temperature
was fitted on validation only. Lower is better for NLL, Brier score and ECE.}
\label{tab:calibration}
\small
\begin{tabular}{llrrr}
\toprule
Split & Score & NLL & Brier & ECE \\
\midrule
Validation & Raw & 0.0685 & 0.0181 & 0.0173 \\
Validation & Temperature-scaled & 0.0653 & 0.0186 & 0.0109 \\
Test & Raw & \RawTestNLL{} & \RawTestBrier{} & \RawTestECE{} \\
Test & Temperature-scaled & \ScaledTestNLL{} & \ScaledTestBrier{} & \ScaledTestECE{} \\
\bottomrule
\end{tabular}
\end{table}

The source-category result was more cautionary than the aggregate AUPRC. At
the unchanged 0.50 boundary, sensitivity was 0.980 for 149 dyed-lifted polyps
but 0.333 for the six standard polyps; their mean scaled scores were 0.947 and
0.319, respectively. The subgroup is too small for a precise standalone
estimate, but the direction and the composition of Table~\ref{tab:composition}
show that classifier performance cannot be generalised to ordinary polyp
detection. All 18 test false positives occurred within the 147
\texttt{dyed-resection-margins} negatives (source-specific false-positive rate
0.122); no false positive occurred in the other negative source categories.
The seven false negatives comprised four standard and three dyed-lifted
polyps. This concentration around dye-associated categories is consistent with
a procedural/source artefact confounder, although the dataset cannot establish
which image feature caused an individual prediction. The finding was exposed
by the protocol's source-labelled export and is retained as a substantive audit
result.

\subsection{RQ1: high-score localisation discordance}

For the primary checkpoint, \PrimaryHighN{} of 1,000 standard-polyp images
had a scaled target score of at least 0.90. On these identical cases, peak
failure was \GradCamDiscordance{} for Grad-CAM,
\GradCamPpDiscordance{} for Grad-CAM++, \XGradCamDiscordance{} for XGrad-CAM,
\HiResCamDiscordance{} for HiResCAM and \EigenCamDiscordance{} for Eigen-CAM
(Table~\ref{tab:primary_audit}). Method choice therefore changed the spatial
interpretation while the classifier output was held fixed. In paired tests,
Grad-CAM++ exceeded Grad-CAM failure by 11.6 percentage points (95\% CI
5.2--18.6; Holm-adjusted $p=0.008$). By contrast, Eigen-CAM and HiResCAM
differed by only 1.7 points (95\% CI $-1.7$ to 5.2; adjusted $p=1.000$), so
their peak-localisation ordering should not be overinterpreted.

\begin{table}[!htbp]
\centering
\caption{Primary spatial audit reported on two explicitly separated
populations. Panel A uses the 172 images with primary-checkpoint
temperature-scaled target score $\geq0.90$; Panel B uses the complete 1,000-image
audit. Every outcome within a panel uses the same images. Values are estimates
with 95\% image-bootstrap percentile intervals (2,000 resamples). Random
heatmaps are single seeded spatial baselines; random deletion uses 20
independent exact-area draws per image. Dashes denote outcomes that do not
apply.}
\label{tab:primary_audit}
\small
\resizebox{\linewidth}{!}{%
\begin{tabular}{lrrrrr}
\toprule
Method & $n$ & Peak failure & CAM--mask IoU & Energy in mask & Logit drop \\
\midrule
\multicolumn{6}{l}{\textit{Panel A: primary high-score population}} \\
Grad-CAM & 172 & .506 (.430--.581) & .219 (.196--.243) & .277 (.246--.308) & .798 (.579--1.020) \\
Grad-CAM++ & 172 & .622 (.552--.692) & .169 (.148--.190) & .213 (.187--.239) & .672 (.475--.870) \\
XGrad-CAM & 172 & .547 (.471--.616) & .202 (.172--.231) & .282 (.242--.320) & .518 (.309--.722) \\
HiResCAM & 172 & .058 (.023--.093) & .350 (.332--.368) & .482 (.458--.505) & 1.657 (1.489--1.835) \\
Eigen-CAM* & 172 & .041 (.012--.076) & .452 (.426--.480) & .686 (.654--.716) & 1.687 (1.495--1.885) \\
Random heatmap & 172 & .826 (.767--.878) & .084 (.079--.090) & .154 (.135--.174) & -- \\
Random deletion & 172 & -- & -- & -- & 1.906 (1.697--2.111) \\
Centre baseline & 172 & .471 (.395--.547) & .283 (.259--.307) & .341 (.307--.376) & -- \\
\addlinespace
\multicolumn{6}{l}{\textit{Panel B: complete spatial-audit population}} \\
Grad-CAM & 1,000 & .764 (.738--.790) & .126 (.118--.134) & .173 (.161--.184) & $-.359$ ($-.442$--$-.272$) \\
Grad-CAM++ & 1,000 & .796 (.771--.820) & .130 (.122--.138) & .178 (.169--.189) & $-.370$ ($-.458$--$-.291$) \\
XGrad-CAM & 1,000 & .722 (.694--.749) & .134 (.125--.144) & .197 (.185--.210) & $-.600$ ($-.679$--$-.521$) \\
HiResCAM & 1,000 & .402 (.372--.431) & .219 (.210--.228) & .316 (.304--.328) & .321 (.243--.399) \\
Eigen-CAM* & 1,000 & .454 (.423--.484) & .269 (.255--.283) & .379 (.360--.398) & $-.192$ ($-.288$--$-.095$) \\
Random heatmap & 1,000 & .832 (.809--.855) & .083 (.080--.085) & .154 (.146--.162) & -- \\
Random deletion & 1,000 & -- & -- & -- & .562 (.474--.656) \\
Centre baseline & 1,000 & .519 (.489--.551) & .253 (.243--.264) & .321 (.307--.336) & -- \\
\bottomrule
\end{tabular}
}
\end{table}

\begin{figure}[!htbp]
    \centering
    \includegraphics[width=\linewidth]{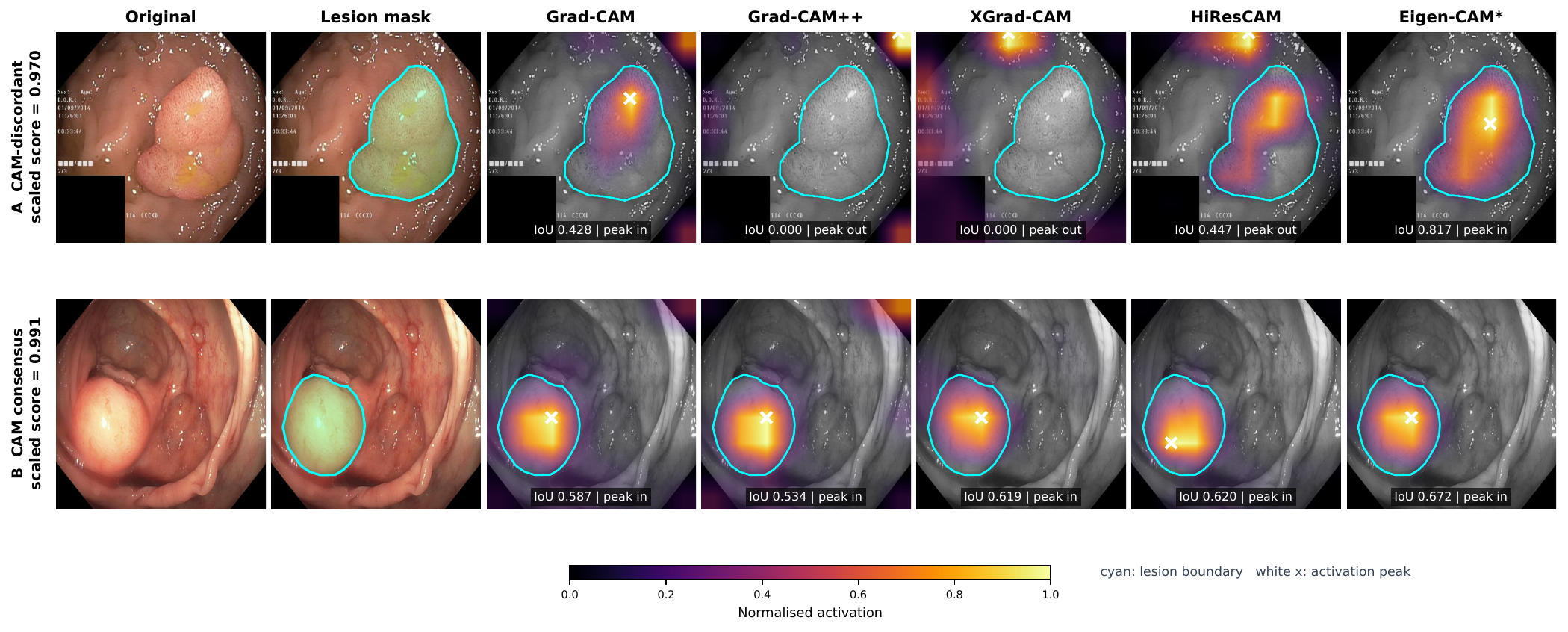}
    \caption{Rule-selected qualitative evidence. Row A is CAM-discordant and
    row B is five-CAM consensus. Columns retain the original image and explicit
    lesion mask. All CAM panels use identical contrast conventions (greyscale
    anatomy, inferno activation, cyan mask, white peak). Rendered metrics are
    checked against the paired export within a 0.002 IoU tolerance. Eigen-CAM
    is class-agnostic. Source images: HyperKvasir, Borgli et al.
    \citep{borgli2020hyperkvasir}, licensed CC BY 4.0.}
    \label{fig:cam_examples}
\end{figure}

Figure~\ref{fig:cam_examples} complements the aggregate results with two
rule-selected examples. The original colour frame and mask are
shown before the five CAMs. Each heatmap uses the same greyscale image base,
inferno activation scale, cyan lesion contour and white activation peak. The
display makes both the activation and its relationship to the mask visible;
it does not substitute for the quantitative audit.

\subsection{RQ2: score, model and spatial-confounder sensitivity}

Raising the temperature-scaled-score threshold from 0.80 to 0.95 reduced the
eligible sample count and did not force convergence between CAMs
(Fig.~\ref{fig:calibration}B). A score threshold changes which images enter the
conditional analysis: the counts were 239, 210, 172 and 121.

Across the nine independently trained checkpoints, each checkpoint's own
$s_i\geq0.90$ subset ranged from 38 to 356 images. On those changing subsets,
the range and ordering of failure rates varied, but method dependence remained
visible (Fig.~\ref{fig:robustness}A). Across all checkpoints, the minimum--maximum conditional failure
ranges were 6.6--97.4\% for Grad-CAM, 11.1--97.0\% for Grad-CAM++,
17.7--60.2\% for XGrad-CAM, 0.0--94.7\% for HiResCAM and 0.0--21.6\% for
Eigen-CAM. The fixed primary 172-image analysis separated checkpoint effects
from this own-score conditioning: corresponding ranges were 10.5--93.0\%,
13.4--98.3\%, 20.3--59.9\%, 2.3--65.7\% and 4.1--25.0\%
(Fig.~\ref{fig:robustness}B). Method rankings were therefore not universal.
Lesion size also mattered differently by method. For example,
centre-baseline failure increased from 18.5\% for large lesions to 76.1\% for
small lesions, whereas primary-checkpoint HiResCAM failure was 9.3\%, 2.8\%
and 6.5\% in the large, medium and small strata, respectively. The adjusted
primary-checkpoint model also included lesion area and centrality
(Table~\ref{tab:adjusted}); its role is to separate obvious geometric
opportunity from residual method differences, not to imply a causal CAM
effect.

\begin{table}[!htbp]
\centering
\caption{Adjusted pointing-failure model for primary-checkpoint CAMs among
scaled scores $\geq0.90$. Odds ratios use Grad-CAM as method reference and
image-cluster bootstrap intervals.}
\label{tab:adjusted}
\small
\begin{tabular}{lrr}
\toprule
Term & Odds ratio & 95\% bootstrap CI \\
\midrule
Grad-CAM++ versus Grad-CAM & 1.628 & 1.264--2.199 \\
XGrad-CAM versus Grad-CAM & 1.180 & 0.782--1.815 \\
HiResCAM versus Grad-CAM & 0.058 & 0.025--0.100 \\
Eigen-CAM* versus Grad-CAM & 0.040 & 0.011--0.074 \\
Lesion area per 0.10 frame fraction & 1.186 & 1.007--1.498 \\
Lesion-centre distance per 0.10 & 1.522 & 1.043--2.274 \\
Scaled score per 0.10 & 0.521 & 0.218--1.196 \\
\bottomrule
\end{tabular}
\end{table}

After adjustment, HiResCAM and Eigen-CAM retained markedly lower failure odds
than Grad-CAM, while Grad-CAM++ had higher odds and the XGrad-CAM interval
included one. Greater centre distance was associated with higher failure odds;
the scaled-score interval also included one. The positive adjusted lesion-area
coefficient differs from the unadjusted centre-baseline strata and reinforces
that these conditional associations should not be read as causal effects.

\subsection{RQ3: localisation and deletion response}

\begin{figure}[!htbp]
    \centering
    \includegraphics[width=\linewidth]{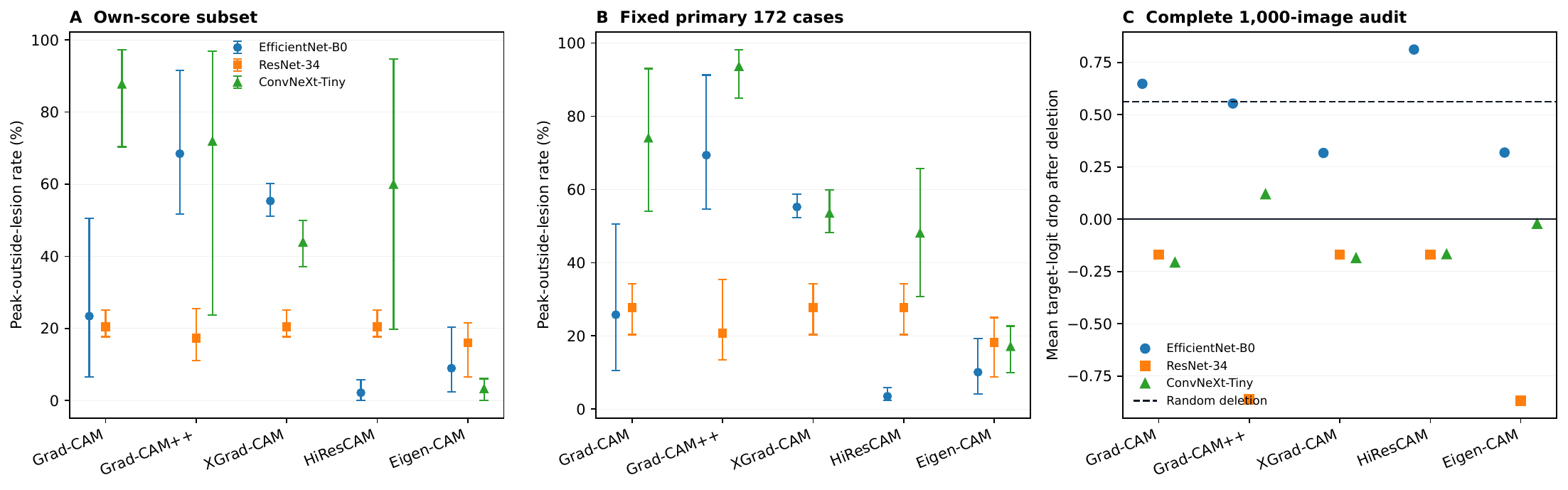}
    \caption{Architecture--seed robustness across all nine checkpoints. Points
    are architecture means and whiskers are across-seed ranges. (A) Peak
    failure on each checkpoint's own score-defined subset ($n=38$--356).
    (B) Peak failure on the fixed 172 images selected by the primary checkpoint.
    (C) Mean target-logit drop over all 1,000 images after replacing the
    top-attributed 20\% with the normalised-image mean; positive values indicate
    supportive evidence under this perturbation. The dashed horizontal line in
    panel C marks the repeated random-deletion mean (0.562). Eigen-CAM is
    class-agnostic.}
    \label{fig:robustness}
\end{figure}

Perturbation response depended on both attribution method and score
conditioning (Fig.~\ref{fig:robustness}C). In the 172-image high-score
population, all five CAMs produced positive mean logit drops, ranging from
0.518 to 1.687, whereas the repeated random-deletion control produced the
largest mean drop, 1.906. Across the complete 1,000-image audit, HiResCAM was
the only CAM with a positive mean logit drop (0.321), while random deletion
remained positive at 0.562 (95\% CI 0.474--0.656). Thus, the sign and magnitude
of the deletion response changed with the analysed population, and a positive
drop alone did not establish attribution specificity.

Spatial overlap and perturbation response also did not rank the CAMs
identically in the complete population. Eigen-CAM had the largest mean IoU
(0.269) but a negative mean logit drop ($-0.192$); Grad-CAM, Grad-CAM++ and
XGrad-CAM also had negative mean drops. An activation map could therefore
overlap the lesion while deletion left the target logit unchanged or increased
it. This is not proof that either metric is intrinsically superior:
overlap asks whether the map resembles a lesion location, while deletion asks
how the trained model reacts to a specific altered input. Their disagreement
is precisely why the framework reports them separately.

\subsection{RQ4: frozen external spatial transfer}

\begin{figure}[!htbp]
    \centering
    \includegraphics[width=\linewidth]{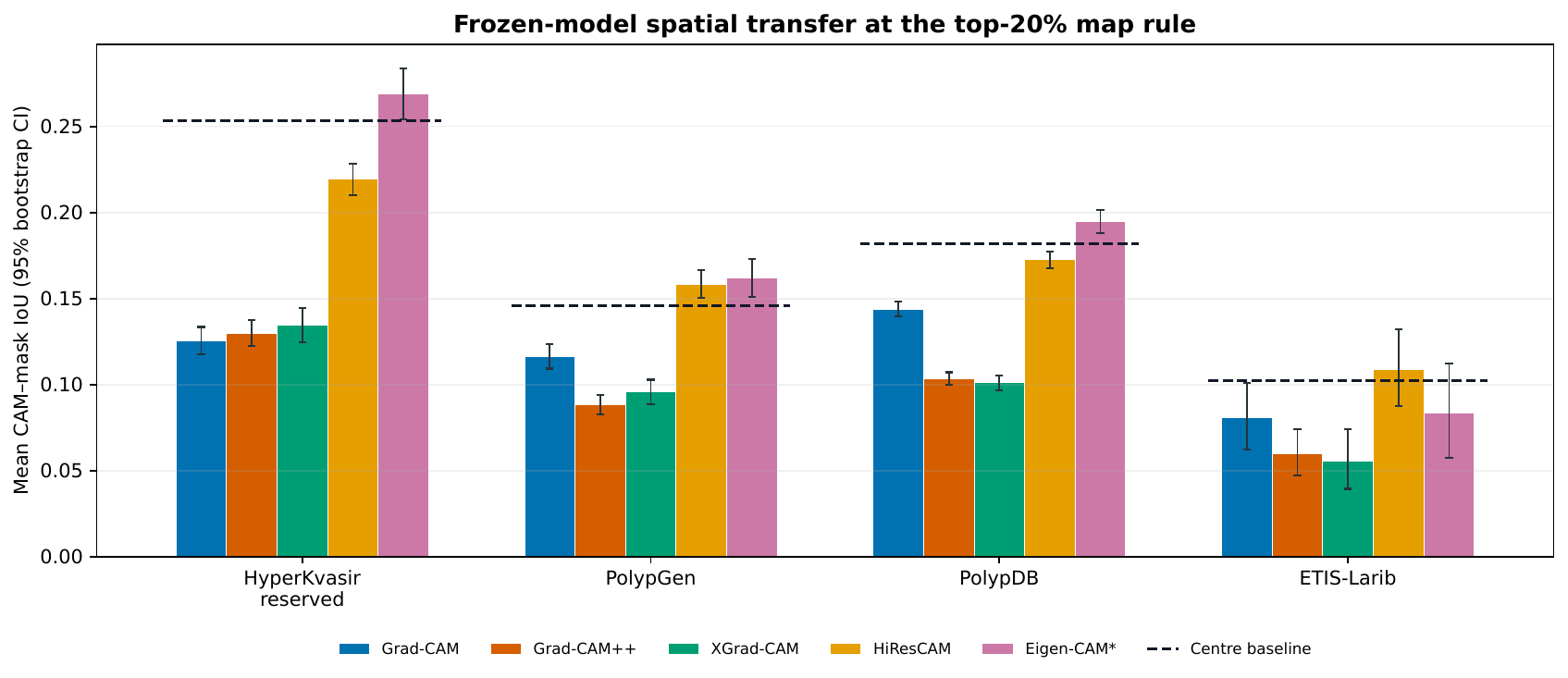}
    \caption{Frozen visible-mask transfer. Bars show mean top-20\%
    CAM--mask IoU with 95\% image-bootstrap intervals for the selected
    checkpoint. The three external cohorts contain no compatible negative
    controls; 125 mask-empty PolypGen frames were excluded from localisation.
    Dashed horizontal segments show the centre-baseline mean for each cohort.
    Eigen-CAM is class-agnostic.}
    \label{fig:transfer}
\end{figure}

The validation-selected model and its temperature were frozen for all three
external cohorts. Mean top-20\% IoU changed by dataset and method
(Fig.~\ref{fig:transfer}). At the scaled 0.90 threshold, eligible counts were
342/1,410 for PolypGen, 748/3,934 for PolypDB and 16/193 for ETIS. Across the
five CAMs, mean IoU ranged from 0.088--0.162, 0.101--0.195 and 0.056--0.109 in
those cohorts, respectively. The centre baseline itself
reached 0.146 in PolypGen and 0.182 in PolypDB, cautioning against interpreting
absolute lesion overlap as a method win. The small high-score ETIS subset
cannot support stable conditional rankings. These results are not
classifier validation; they show that identical frozen code can expose
collection-dependent spatial behaviour.

\section{Discussion}
\label{sec:discussion}

\subsection{Principal findings}

The results answer the four questions through one coherent progression. First,
a high scaled target score did not guarantee that a CAM peak lay in the lesion.
Second, the result depended on method, score threshold, architecture, seed and
lesion geometry rather than on a single arbitrary checkpoint. Third, lesion
overlap and deletion response were empirically distinct. Fourth, a frozen
model's spatial behaviour changed across datasets. endoExplain's contribution
is to make each dependency visible in a reproducible, case-linked record.

Calibration strengthens the wording but does not transform the score into
clinical certainty. Temperature scaling modestly improved test ECE and NLL,
yet Brier score slightly worsened, intermediate reliability bins were sparse,
and only three standard polyps informed validation. The correct term is
therefore ``temperature-scaled target score'', not patient-level probability.
Reporting this nuance prevents the 0.90 boundary from being mistaken for a
clinical operating point.

The source-category audit is equally important. Aggregate discrimination was
dominated by dyed-lifted polyps, whereas the spatial and external cohorts
contained standard polyps. This makes the striking fall in standard-polyp score
an expected distribution-shift warning and prevents a broad polyp-detector
claim. The observation does not invalidate the audit: it illustrates why
model-level AUPRC, temperature-scaled score and local evidence should be traceable to
their source categories. A less transparent workflow could have reported high
test AUPRC and plausible heatmaps without revealing that they evaluated
different positive compositions. The concentration of false positives in
dyed resection margins further suggests that the collapsed binary target can
reward procedure-associated appearance rather than the intended general polyp
construct; this is an association, not a causal feature attribution.

\subsection{Interpretation of CAM differences}

No method should be declared the universal winner from lesion IoU alone.
Eigen-CAM can align with a conspicuous lesion while remaining class-agnostic.
Centre agreement can be inflated when lesions are central. Gradients can
highlight discriminative boundaries or context rather than full lesion extent.
HiResCAM's theoretical guarantee applies to model output decomposition under
specified architectures; it does not guarantee agreement with a human lesion
mask, immunity to preprocessing or a favourable deletion intervention.
Consequently, method labels, target definition and metric semantics must
remain visible wherever a heatmap is shown.

The qualitative figure is valuable precisely because the colour design no
longer conceals activation. A colour image under a translucent jet map can
make weak or widespread activation difficult to distinguish. Greyscale anatomy
with a perceptually ordered inferno overlay increases activation contrast,
while a cyan contour and white maximum separate annotation from attribution.
Still, attractive contrast is a communication choice, not a validation result;
the case-selection rule and numerical labels guard against visual cherry
picking.

Deletion adds a second axis of evidence but has its own failure modes. Replacing
pixels by a mean value can introduce unnatural boundaries, discard negative
evidence or alter texture statistics. Negative logit drops should therefore be
read as failed support under this particular stress test, not proof that the
model never used the region. The random-deletion control produced positive
drops in both populations and exceeded all CAMs in the high-score population,
so a positive deletion response cannot establish attribution specificity. More
comprehensive future audits can add insertion,
AOPC, ROAD and retraining-based tests, at considerably greater computational
cost \citep{petsiuk2018rise,hooker2019roar,rong2022road}.

\subsection{Methodological and practical implications}

For research reporting, the minimum defensible unit is a row linking image,
model, class score, temperature-scaling status, CAM method, mask metric and perturbation
outcome. Aggregate classifier and explanation tables can then be derived
without losing case correspondence. Content hashes should define roles when
the same public image can appear in labelled and segmented distributions.
Architecture and seed repeats should accompany claims about an explanation
method, because a CAM operates on a learned representation, not only on an
architecture name.

For software interfaces, score and heatmap should not be fused into one
assurance cue. A review display can identify the target and CAM, indicate
whether temperature scaling was applied, expose the spatial threshold and
signal disagreement with a reference mask when one exists. endoExplain supplies the audit plumbing for
such research workflows, but this paper does not evaluate reader behaviour,
automation bias, alert fatigue or clinical decision benefit.

\subsection{Limitations}

This retrospective study uses public image datasets and no prospective or
multicentre local cohort. The classifier target is composition-confounded and
does not establish clinical polyp detection. The standard-polyp test subgroup
contains only six images, precluding a precise subgroup performance estimate.
The three external spatial cohorts lack compatible negative controls, so they
cannot estimate external discrimination, specificity or calibration. ETIS is
small after duplicate-frame union, and PolypGen excludes frames without a
visible annotated lesion from localisation endpoints.

Patient and sequence identifiers were unavailable for the analysed exports;
source-centre labels were retained only where distributed for PolypGen and
PolypDB. Images from the same procedure may therefore violate independence and
make image-bootstrap intervals too narrow. The lesion mask is treated as a
spatial reference but does not encode every context feature legitimately used
by a classifier. Pointing game favours any single in-mask maximum, while IoU
depends on retained map area; neither is a causal explanation metric. The
adjusted logistic analysis can reduce measured geometric confounding but cannot
remove unmeasured acquisition or source-category confounding.

Temperature scaling was fitted to the imbalanced mixed target and is only
one-dimensional. ECE depends on binning and can appear small in strongly
bimodal predictions. The deletion intervention can be out of distribution.
Only CAM-style maps and two baselines were evaluated; perturbation-heavy or
concept-based explanations may behave differently. Finally, the visual display
has not undergone a reader study, and no clinical-use or medical-device claim
is made.

Future work should use a patient-grouped multicentre cohort with standard
polyps and compatible negatives, preregister clinical operating points, and
fit calibration on a representative validation population. Sequence-aware
bootstrap or hierarchical modelling should replace image resampling when
identifiers are available. A broader faithfulness battery and parameter
randomisation can be added, followed by a separately powered reader study.

\section{Conclusions}
\label{sec:conclusion}

A high score, a lesion-aligned CAM and a positive perturbation response are
different observations. endoExplain turns their relationship into a
content-audited, calibration-aware, paired and reproducible protocol across methods,
models and datasets. In this study the protocol not only identified
score--localisation discordance, but also exposed a source-category confounder
that sharply limits classifier claims. The defensible outcome is therefore an
auditable evidence record, not a claim that a visually plausible heatmap proves
where or why a colonoscopy classifier decided.

\section*{Acknowledgements and declarations}

\paragraph{Acknowledgements.} The authors thank the creators and curators of
the public datasets used in this research.

\paragraph{Ethics statement.} This computational study reanalysed de-identified
public datasets and did not recruit participants, perform an intervention or
access identifiable clinical records. No new human-subject data were
collected. Ethical approval and consent procedures for the source datasets are
reported by their original providers.

\paragraph{Funding.} This research did not receive any specific grant from
funding agencies in the public, commercial or not-for-profit sectors.

\paragraph{Competing interests.} The authors declare no competing interests.

\paragraph{Data availability.} HyperKvasir, ETIS-LaribPolypDB, PolypGen,
PolypDB and LDPolypVideo remain subject to their providers' access conditions
and licences. The locally distributed HyperKvasir licence is CC BY 4.0; Figure
\ref{fig:cam_examples} reproduces two attributed rule-selected examples under
that licence. No dataset archive or source-image collection is redistributed
by this project. The manifest contains hashes and identifiers but no image
data. The analysed PolypGen package contains a CC BY-NC-SA 4.0 licence file,
whereas its README and data descriptor describe CC BY 4.0. We followed the
stricter terms: non-commercial research use, attribution, no source-image
redistribution and aggregate reporting. The public PolypDB OSF record expressly
opens the collection for research and education but sets no standardised node
licence; our use is limited accordingly to research analysis and aggregate
reporting, with no redistribution. CVC-ColonDB and CVC-ClinicDB were excluded
because the required documented approval and signed agreements were unavailable.

\paragraph{Code availability.} Source code, configurations and tests are
available at \url{https://github.com/eroberam/endoExplain}. The archived
software release is available at
\url{https://doi.org/10.5281/zenodo.21508564}. Local datasets, weights and
prediction outputs are excluded from Git.

\paragraph{Declaration of generative AI and AI-assisted technologies in the
manuscript preparation process.} During preparation of this work, the authors
used OpenAI Codex solely to assist with code auditing.

\section*{CRediT authorship contribution statement}

\noindent\textbf{Roberto Alcaraz Machado:} Conceptualization, Methodology,
Software, Data curation, Formal analysis, Investigation, Visualization,
Writing--original draft, Writing--review and editing.

\noindent\textbf{Ana Luc\'ia Rodr\'iguez Blanco:} Validation, Investigation,
Writing--review and editing.

\end{document}